# Synthesis and physical properties of uranium thin-film hydrides UH$_2$ and β-UH$_3$

Evgenia A. Tereshina-Chitrova[1,2], Ladislav Havela[2], Mykhaylo Paukov[2,3], Oleksandra Koloskova[2], Lukas Horak[2], Milan Dopita[2], Mayerling Martinez Celis[2], Miroslav Cieslar[2], Zbynek Šoban[1], Thomas Gouder[4], Frank Huber[4]

[1]*Institute of Physics, Czech Academy of Sciences, 18121 Prague, Czech Republic*
[2]*Faculty of Mathematics and Physics, Charles University, 12116 Prague, Czech Republic*
[3]*Nuclear Fuel Cycle Department, Research Centre Rez Ltd., 25068 Husinec-Rez, Czech Republic*
[4]*European Commission, Joint Research Centre (JRC), Postfach 2340, DE-76125 Karlsruhe, Germany*

**Abstract**

Formation of thin uranium hydrides films, UH$_2$ and β-UH$_3$, synthesized by a reactive dc sputtering of uranium metal, was explored using variable deposition conditions. Obtained stable oxygen-free hydride films were studied by a variety of methods, both in situ (photoelectron spectroscopy - XPS), and ex-situ (x-ray diffraction - XRD, transmission electron microscopy - TEM), electrical resistivity, and magnetometry). Both types of hydrides are ferromagnetic, the Curie temperatures of UH$_2$ and β-UH$_3$ are approx. 120 and 170 K, respectively. Ferromagnetism in the thin films is robust and does not depend on structure details while electrical resistivity data reflect disorder in both types of hydrides.

## I. Introduction

Synthesis of materials in a thin film form affected by interaction with a substrate together with strains imposed by the kinetics of deposition, and by other deposition details such as composition of the sputter gas, provide factors driving materials far from thermodynamic equilibrium. In extreme cases new structure types can be obtained [1]. The interplay of thermodynamic and kinetic factors determines the stability of the phase while structural coherence to the substrate facilitates the growth process. In the case of hazardous substances, such as e.g., pyrophoric uranium hydride, preparation of thin films has several advantages. Well-defined films are ideal objects as they contain very little amount of radioactive material, they are stable [2] and do not self-ignite in air and allow *in-situ* surface science studies, which is of considerable relevance in nuclear energy research. Strong reactivity of uranium with hydrogen and its abundancy makes it valuable as a storage medium for tritium.

From bulk studies of the U hydrides [3] it was known that $UH_3$ has two allotropes, namely the transient α-$UH_3$ phase and β-$UH_3$, into which the α-$UH_3$ phase transforms fast. Both tri-hydrides have cubic unit cells, corresponding to a *bcc* U lattice for α-$UH_3$ with $a$ = 416 pm and a more complex structure for β-$UH_3$ (two different U sites, 8 formula units) with $a$ = 664 pm [3,4]. Available information on α-$UH_3$ suggests that its magnetic properties are very similar to β-$UH_3$ [4]. The latter is a ferromagnet with the bulk Curie temperature $T_C$ = 170 K. The ferromagnetism in thin β-$UH_3$ films was shown to be very robust with similar to the bulk $T_C$ of 178 K [2,5,6]. This is rather surprising as the 5$f$ magnetism in U-systems is very sensitive to structure details and reduced grain size often leads to a reduction of the ordering temperatures [7].

Recently thin film synthesis by reactive sputtering also allowed stabilization of a lower U hydride $UH_2$ [8], which does not form in a bulk form even though other actinides can form stable di-hydrides. To enable the growth of $UH_2$, Havela at al. [8] used a substrate (Si) with a lattice parameter close to that of other stable actinide dihydrides, such as $NpH_2$ and $PuH_2$. Basic magnetic characterization showed that the thin film of $UH_2$ is also ferromagnetic with the ordering temperature $T_C$ ≈ 120 K. However, it was not clear which factors enable the formation of the $UH_2$ films and under which conditions the U sputtering yields the common variant β-$UH_3$.

At a first glance, the difference in the ordering temperatures of β-$UH_3$ and $UH_2$ seems reasonable due to a higher H concentration in the former hydride since hydrogenation (often regarded as a tool to boost the volume [9]) influences the size of individual magnetic moments and strength of exchange interactions in the compounds, determining the ordering temperatures [10]. Closer inspection of the U-U distances – a parameter critical for uranium-based compounds, shows that $d_{U-U}$ = 331 pm in β-UH$_3$ is below the approximate boundary value of the spacing, corresponding to the critical 5$f$-5$f$ wave functions overlap (so-called Hill limit, $d_{U-U}$ = 340-360 pm [11]) so this compound should not be magnetic at all. On the contrary, $d_{U-U}$ in UH$_2$ (378 pm), significantly exceeds the Hill limit, which in principle may give a higher $T_C$ [8–10]. To prove that the lower $T_C$ of UH$_2$ as compared to β-UH$_3$ is its intrinsic feature, i.e., that is not attributed to finite size effects and variations of the microstructure and to test the sturdiness of the ferromagnetic state in the uranium hydrides thin films, systematic study of conditions under which the stable β-UH$_3$ and metastable UH$_2$ tend to grow is required.

In this work, we reveal the structure variabilities in these materials and their relation to the electronic, magnetic and transport properties. In particular, we describe the conditions (temperature, deposition rate, gas pressure, and substrate choice) for the U-H films growth and investigate the samples in-situ (by x-ray photoelectron spectroscopy (XPS) on freshly synthesized surfaces without breaking the vacuum) and ex-situ by micro- (XRD, TEM) and macroscopic (magnetization, electrical resistivity) methods. When available, properties of the films are compared with available data obtained on monolithic bulk materials, coming from experiments with specific transition metals substituting for U. In the films, parameters as texture or residual strain are evaluated in addition to common parameters of the bulk data (lattice parameter, grain size). To avoid decomposition of the hydrides at elevated temperatures, we restrict ourselves to preparing the U-H films at low temperatures (room temperature and below).

## II. Experimental

Thin film samples of uranium hydrides with less than 1 at.% C and O were prepared by reactive sputter deposition in an Ar-H$_2$ atmosphere using a miniature U target (natural uranium, 99.9 wt.% purity). The working gas (mixture of Ar and hydrogen, purified by the Oxisorb® cartridge) with a pressure of 0.5-0.8 Pa contained about 10 % of H$_2$. Details of the home-built dc sputtering equipment can be found elsewhere [5]. We used polished fused silica substrates (amorphous SiO$_2$) and Si wafers with the (001) orientation to prepare the films. Prior to samples preparation, SiO$_2$ substrates were cleaned by annealing at $T$ = 673 K for 1200 s in the chamber with base pressure of 10$^{-9}$ Pa. The Si wafers were cleaned by Ar ion bombardment at $T$ = 523 K.

One of the key deposition parameters was the substrate temperature. Throughout the study, we experimented with cooling the substrates by LN$_2$ to various temperatures down to 77 K. When depositing at room temperature, we used two different approaches. One was to cool the substrate by a flow of gas to room temperature (RT). And the second approach was performed without such cooling. Then the substrate temperature increased to ≈ 350 K by the heat coming from the sputter source (these samples are denoted as self-heated throughout the text of the article). In such a case, short deposition times up to 1000 sec were used to prevent excessive surface heating and hydrogen escape. The characteristic deposition time was chosen experimentally by monitoring XPS spectra of the films (Fig. 1). XPS study was performed using the monochromated Al–K$_α$ radiation with the photon energy h$\nu$ = 1486.6 eV and SPECS PHOIBOS 150 MCD-9 electron energy analyser. Standard correction for transmission function of the electron energy analyser was performed and Shirley secondary electron background was subtracted.

Table 1 reveals sputter deposition parameters used for the thin film synthesis. It also shows whether buffer or a capping Mo layer (or both) were used. Samples for the subsequent resistivity study were neither buffered nor capped, allowing them to oxidize naturally to form a non-conducting UO$_2$ layer on the top. The XPS analysis of deposited films was performed *in situ*; for the capped films it was done before they were covered by a cap. The U target voltage -690 V was used typically for the hydrides deposition. Besides the target voltage one can also vary the target current (see Table 1), tuned by emission parameters from a negatively charged heated filament, contributing to the plasma stabilization.

The structure of the films was explored using x-ray diffraction (XRD) methods on a Rigaku SmartLab diffractometer equipped with a 9 kW copper rotating anode x-ray source and Panalytical X'Pert Pro MRD and Panalytical X'Pert Pro MPD diffractometers with Cu x-ray radiation (wavelength $\lambda$ = 0.15418 nm). Various geometries were used in the experiments. Parallel beam (PB) geometry was used for the x-ray reflectivity (XRR) and grazing incidence angle x-ray diffraction (GIXRD) with the incidence angles of the primary beam, $α_i$ = 0.3 - 10°. For selected samples the obtained diffraction patterns were fitted using the whole powder pattern refinement method (Rietveld method). The computer program MStruct [12] was used for the fitting. Pole figures were collected in the PB setup utilizing the in-plane arm that allows moving the detector to the arbitrary equatorial and axial position [13]. Symmetrical $\theta$-2$\theta$ scans were measured in the Bragg-Brentano geometry.

Table 1. Sputtering details of individual thin films.

| Sample code | Substrate | Substrate temperature (K) | Buffer/Capping | Current (mA) | Deposition time (sec) |
|---|---|---|---|---|---|
| SO1 | SiO$_2$ | RT | -/- | 1 | 3600 |
| SO2 | SiO$_2$ | RT | -/- | 0.7 | 4000 |
| SO3 | SiO$_2$ | RT | -/- | 2.2 | 3000 |
| SO4 | SiO$_2$ | RT | -/- | 1.5 | 3000 |
| SO5 | SiO$_2$ | RT | Mo/Mo | 1.7 | 5000 |
| SO6 | SiO$_2$ | 170 | Mo/Mo | 2.3 | 4000 |
| SO7 | SiO$_2$ | 350 | -/- | 2.2 | 1000 |
| S8 | Si | RT | -/Mo | 2.3 | 4000 |
| S9 | Si | 170 | Mo/Mo | 2.8 | 4000 |
| S10 | Si | 150 | -/Mo | 2.3 | 4000 |

Transmission electron microscopy (TEM) study, allowing for the film thickness check and determination of elemental and phase composition, was performed by means of the JEOL JEM-2200FS microscope operating at 200 kV. The TEM samples were prepared for the studies of cross-section in a scanning electron microscope (SEM) ZEISS Auriga Compact equipped with a focused ion beam (FIB) and easy-lift manipulator designed for In-Situ Lift-Out thin lamella preparation.

Magnetization measurements were conducted using a PPMS 9 (Quantum Design, USA) installation equipped with a VSM to improve sensitivity. The electrical resistivity $\rho(T)$ of the thin film samples was studied by the Van der Pauw method using 30 μm-thick Al wires with 3% of Si, forming contacts to the films by a wire bonder. External magnetic field was applied along the sample surface in both the magnetization and transport studies.

## III. Results and discussion
### 3.1. XPS study

Overview XPS spectra of each sample (not shown here) were used for the basic check of chemical composition and absence of impurities (O, C). The H-1$s$ states are part of the valence band of hydrides and cannot be discriminated in the photoelectron spectra, precluding quantification of the H concentration. Nevertheless, the hydride formation is directly reflected in changes of shape of the U-4*f* core-level spectra (Fig. 1). Starting from pure uranium, the spectra vary continuously when increasing partial $H_2$ pressure in the sputter gas (Ar). The formation of a homogeneous hydride can be concluded when the spectra no longer change upon the $H_2$ pressure increase. This happens typically around 5% of $H_2$ in Ar (and we safely work with 10% of $H_2$ in Ar).

The most commonly studied spectral lines, U-4*f*, are shown in Fig. 1, which compares the spectra of pure U metal film with those of the U-H samples deposited at room temperature (without cooling the substrate, in other words, self-heated samples) for times shorter and longer by about a factor of 2 than the typical time of 1000 sec. In the latter case, the hydrogen concentration was most likely lower (intensity of the peaks starts to increase towards the level of U) due to excessive heating of the surface by the sputter source, although it is still more similar to the full hydride. The peak shapes and intensities of the three samples are noticeably different while the maxima shift only very little, by 0.2 eV. The shift of peaks towards the higher binding energies in the hydrides as compared to pure U metal is generally attributed [5] to lowering of the correlation energy between the 4*f* shell and outer electrons, transferred from the U atom due to the chemical bonding. High-intensity asymmetric 4*f* lines of U metal become broader in the hydrides due to the crystal lattice expansion, causing narrowing of the 5*f* band and increase of the quasiparticle density of states at the Fermi level. This fact can be illustrated by the increase of the Sommerfeld coefficient of electronic specific heat, reaching ≈ 30 mJ/mol $K^2$ [14], i.e. about 3 times as much as in α-U. The almost twofold decrease of maximum intensities can be also affected by the lower density of hydrides. All these spectral features are present in both $UH_2$ and $UH_3$ [5,8] so at this point we cannot say exactly what kind of hydride has been formed and we can only register its formation. (Later, we identified the spectra of self-heated samples as containing predominantly the $UH_2$ phase.)

Besides the 4*f* lines, we studied also the respective 5*d* (Fig. 2) and 6*p* (Fig. 3) lines. They generally have much lower intensity and hence they are not studied routinely for U compounds, however, due to the much larger spatial extent of the respective electronic states they should be more sensitive to the changes of occupation of extended valence states (6*d*, 7*s*). Indeed, we see that the shift between U and its hydrides is 0.6 eV for 5*d* and over 1 eV for the 6$p_{3/2}$ peak. In all the cases, UH$_2$ and UH$_3$ are practically identical as to their energy and shape, while there is certain difference in intensities (Figs. 2 and 3), in analogy with the 4$f$ spectra.

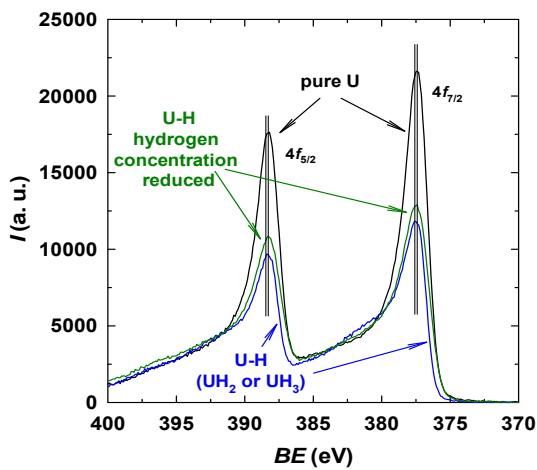

Fig. 1. U-4$f$ core level spectra of U (black) and U-H hydrides. For the latter we show the spectrum of a formed hydride (UH$_2$ or UH$_3$, in blue) and the U-H spectrum with the hydrogen concentration reduced by exposure to elevated temperatures (green). Each spectrum consists of the 4$f_{5/2}$ and 4$f_{7/2}$ peaks, split by 11 eV due to spin-orbit interaction.

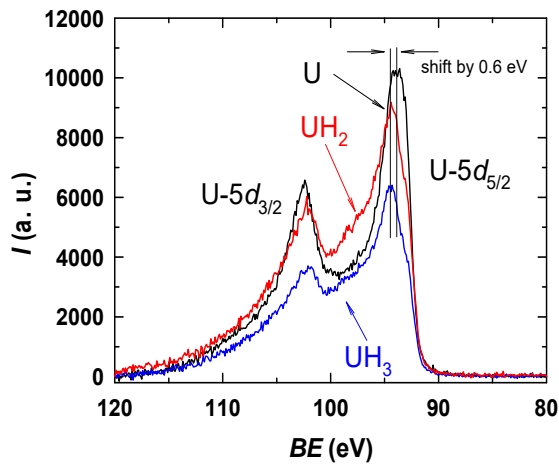

Fig. 2. U-5$d$ core level spectra of U, UH$_2$, and UH$_3$. Each spectrum consists of the 5$d_{3/2}$ and 5$d_{5/2}$ peaks, separated by 8 eV due to spin-orbit interaction.

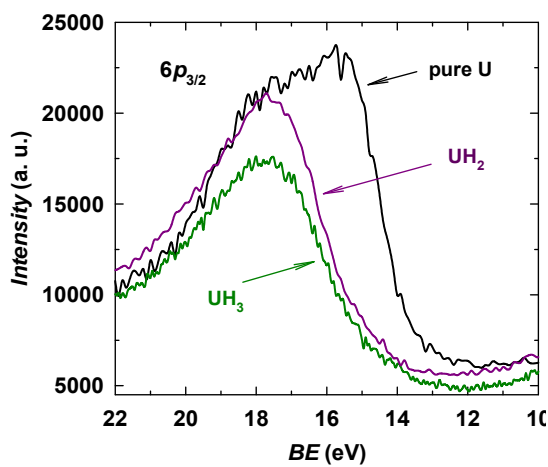

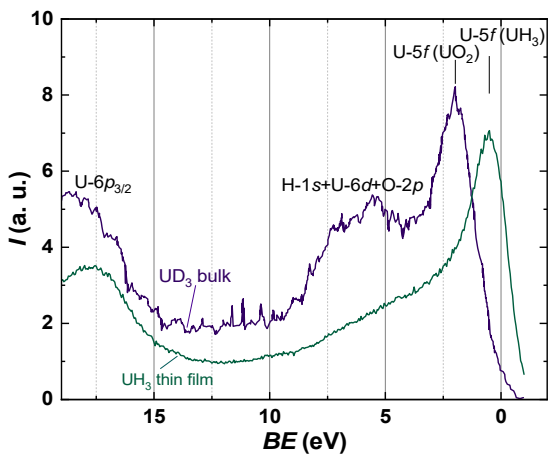

Fig. 3. U-6$p_{3/2}$ spectra for U, UH$_2$ and UH$_3$.

Fig. 4. Valence-band XPS spectrum of the UH$_3$ thin film also contains the U-6$p_{3/2}$ emission. The spectrum is compared with historical bulk UD$_3$ data [14].

The advantage of using sputter deposited films and in-situ electron spectroscopy is well illustrated on valence-band XPS data (Fig. 4), comparing the spectrum of a pure UH$_3$ film with earlier published data on bulk UD$_3$ having a freshly cleaved surface [14]. The comparison of data suggests that the spectrum in Ref. [14] was contaminated by more than 50% of UO$_2$ present at the surface, manifested by the maximum at 2 eV BE due to $5f^{\,2}$ state, while the O-2$p$ states contribute around 6 eV BE. Hence, the 5$f$ states appear practically at the Fermi level in U hydrides (still visible as a minor spectral feature for Ref. [14] data). From the XPS data presented, we can certainly conclude the purity of the sample and formation of the hydride phase, while the exact composition of the hydride cannot be determined unambiguously and can only be revealed using a *combined* XRD (complimented with TEM studies) and magnetometry data, which we present below.

### 3.2. XRD characterization of the samples
*3.2.1. Room temperature deposition on amorphous SiO$_2$ substrates*

Figure 5 compares GIXRD profiles of selected samples deposited on glass substrates at room temperature (gas cooling of the substrate) using various deposition currents (see Table 1). The patterns look very similar being different only in fine details around the peaks in the vicinity of 30 deg. Despite the fact that UH$_2$ and UH$_3$ have very different crystal structures and lattice parameters, they are practically indistinguishable in the presented data. Broadening of diffraction lines, caused by combination of microstrain and limited grain size down to several nm, together with very strong texture which often appears in the samples deposited on glass substrates [5,6] restricts the abilities of the XRD method. The overlap of several strong peaks contributes to uncertainty.

Closer inspection of diffraction patterns shows that the deposition on the room temperature (cooled) substrate results in a mixture of β-UH$_3$ (prevailing) and UH$_2$ phases (Fig. 5). Lower deposition current (slower reaction with hydrogen) increases the amount of the UH$_2$ phase at the expense of the β-UH₃ phase. But the latter is still present in the samples even with the current as low as 0.7 mA (sample SO2, see the magnetization section). Deposition current naturally influences thickness of the layers (Fig. 6); the lower the current, the fewer material is deposited. Figure 6 provides total thicknesses of the films as we cannot reasonably separate for the fitting UH₃ and UO₂ layers having very close densities and rough interface as oxygen penetrates the U-H layer unevenly.

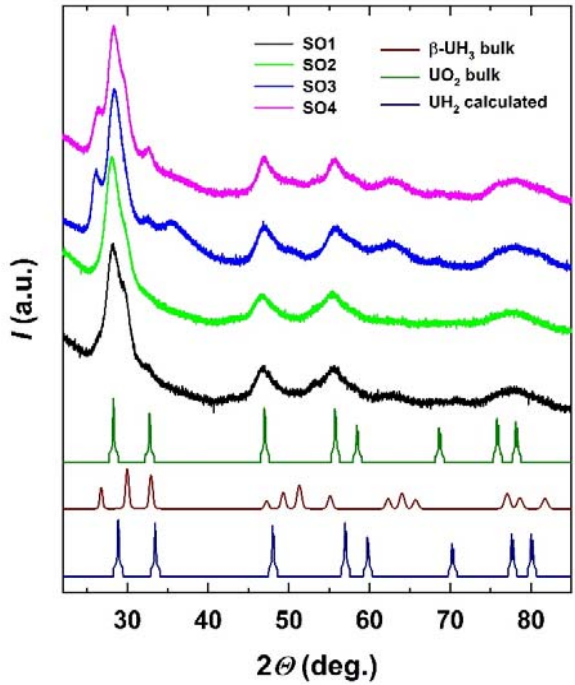

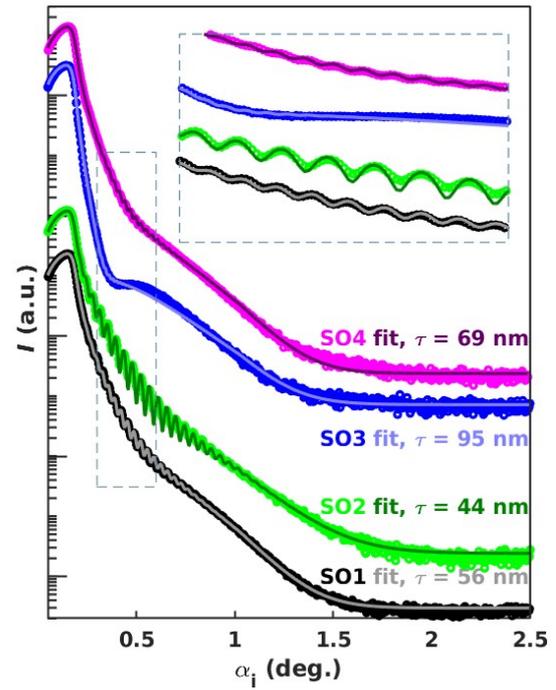

Fig. 5. The GIXRD patterns of the uncapped U-H films prepared with air cooling at room temperature on the SiO₂ substrates measured with an angle of incidence of a primary beam $\alpha_i = 1.5°$. For comparison, XRD data for bulk UO₂ and β-UH₃ are shown. The spectrum of UH₂ was calculated using the lattice parameter found in Ref. [8].

Fig. 6. X-ray reflectivity curves of the U-H films prepared with air cooling at room temperature on the SiO₂ substrates. Fits to the experimental data and calculated thicknesses of the layers $\tau$ are shown.

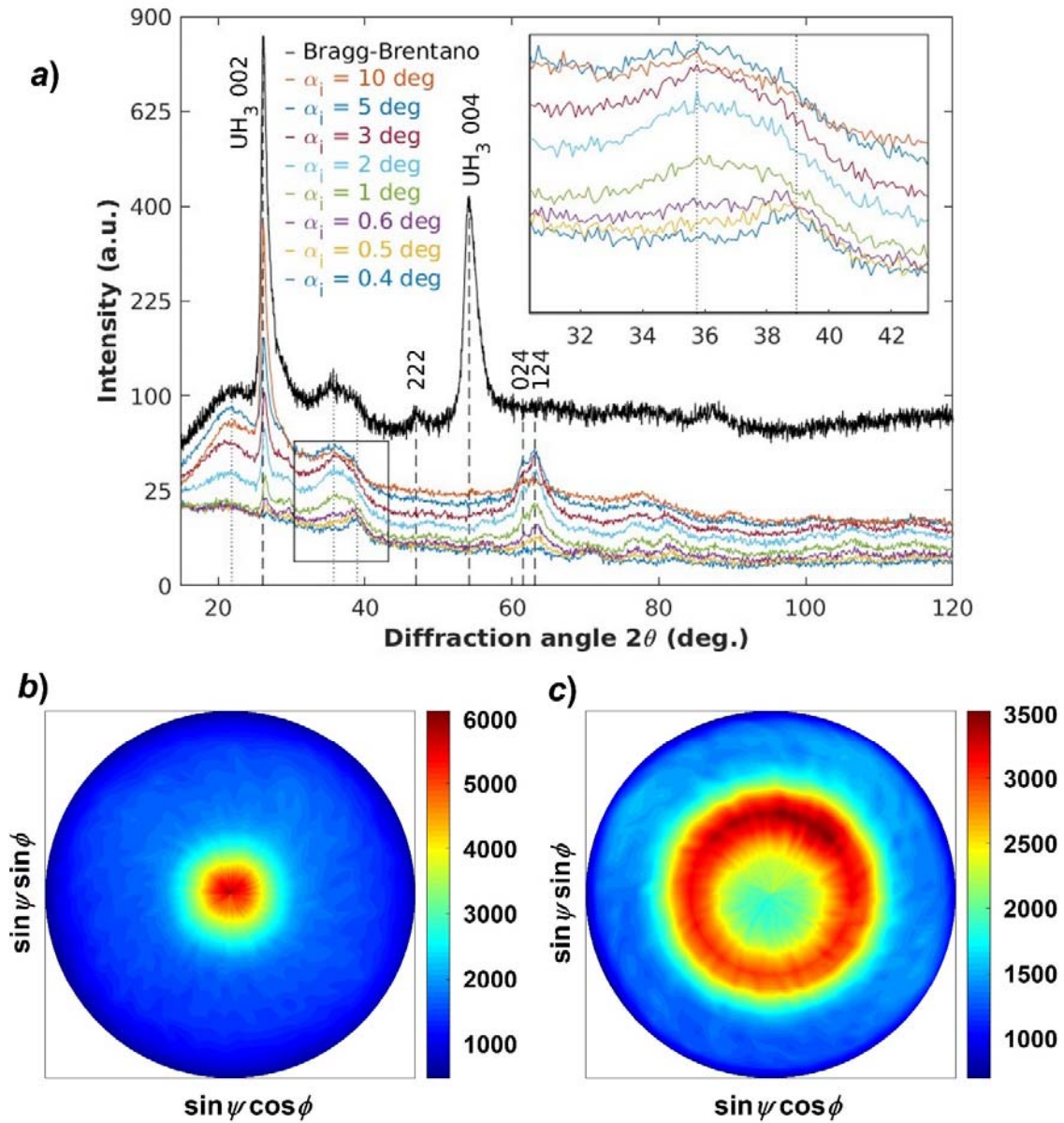

Fig. 7. (*a*) Comparison of the GIXRD patterns of a Mo-capped β-UH₃ film (sample SO5) measured with various angles of incidence of the primary beam ($α_i$ = 0.4-10°) with the data obtained in the Bragg-Brentano geometry. Inset shows enlarged area around spurious peaks at 35.8°, 38.8°, which can be associated with a small amount of non-hydrogenated U or U-Mo. (*b*) Experimental (002) and (210) (*c*) pole figures of the β-UH₃ thin film sample.

An additional aspect making the phase identification more difficult is a pronounced texture, which can dramatically reduce the number of observable diffraction peaks. For β-UH$_3$, the texture is typically of the (00$l$) type and is best observed in the Mo-capped samples, i.e. when the interaction with air and moisture is minimized. Figure 7 (*a*) compares the GIXRD profiles of the Mo-capped UH$_3$ film (sample SO5 in Table 1) measured with several angles of the incident beam $\alpha_i$ and data obtained in Bragg-Brentano geometry. (The increase of $\alpha_i$ increases penetration depth, being nearly independent of the diffraction angle). In the Bragg-Brentano measurement, only the (002) and (004) peaks of β-UH$_3$ are observed, suggesting a pronounced (00$l$) texture of the sample. Pole figure measurements (Fig. 7, *b* and *c*) confirm the (00$l$) texture, i.e. the (00$l$) lattice planes are oriented predominantly parallel to the sample surface while there is a random orientation of crystallites within the sample plane. Interestingly enough, for oxide thin films [15], the stabilizing effect was observed only if epitaxial growth was induced. This is obviously not the case of the uranium hydrides, where the samples grow with strong preferential orientation but have a misorientation within the basal plane.

As a rule, we detect a sizeable compressive residual stress in the samples. For the sample SO5 in Fig. 7 the out-of-plane parameter is *a* = 683 pm, the in-plane parameter is *a* = 664 pm, and the determined stress-free lattice parameter *a* = 681.6(3) pm is noticeably higher than that of bulk β-UH$_3$, *a* = 664.4 pm [3]. Large strain usually slows down corrosion in air by preventing formation of cracks enabling propagation of oxidation into the bulk of the film. As a result, the body of the films remains metallic on the timescale of years.

*3.2.2. Influence of temperature and type of substrate*

We find that in the samples prepared on a glass substrate at room temperature with no air cooling (self-heated to ≈350 K during deposition) the UH$_2$ phase prevails irrespective of the deposition current (Fig. 8, bottom). Figure 8 compares the XRD patterns of the UH$_2$ samples on the SiO$_2$ substrates obtained at temperatures as low as 170 K (top) and 350 K (bottom). UH$_2$ deposited at low temperatures has crystallites of 16 nm (Fig. 8, top), while the mean crystallite size surprisingly drops to 3-4 nm in films deposited at 350 K (Fig. 8, bottom). The structure refinement using the MStruct program [12] confirmed the cubic *fcc* structure in the UH$_2$ sample prepared at *T* = 170 K with the lattice parameter *a* = 533.17 ± 0.07 pm. The film exhibited a substantial compressive residual stress of σ = −1.61 ± 0.05 GPa.

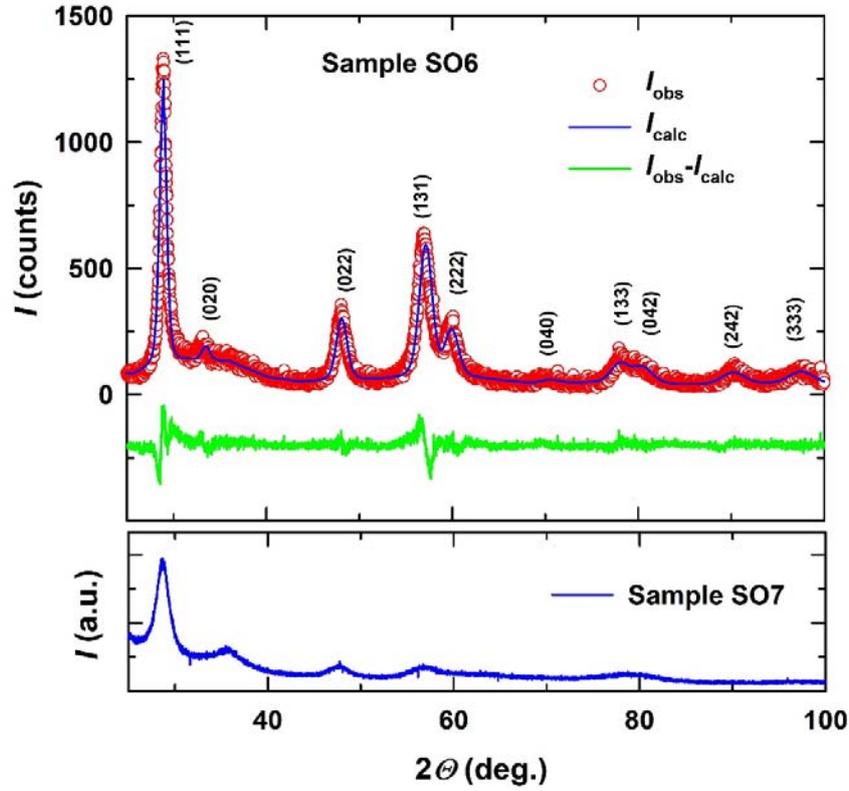

Fig. 8. Top: GIXRD pattern of the UH$_2$ thin film prepared at 170 K on a SiO$_2$ substrate obtained with an angle of incidence of the primary beam $\alpha_i$ = 1.5°. Bottom: The XRD data for a self-heated to 350 K UH$_2$ thin film sample on a SiO$_2$ substrate.

By using suitable substrates, we can further facilitate formation of the UH$_2$ phase at all temperatures. For instance, Si has a cubic structure with a lattice parameter $a$ = 543.1 pm, which is close to that of NpH$_2$ (534.3 pm) and PuH$_2$ (535.9 pm) [16,17]. Figure 9 compares XRD diffraction patterns of UH$_2$ films on Si (100) substrates obtained at room temperature and below. UH$_2$ with crystallites mean size reaching almost 100 nm is obtained using a Si substrate at 173 K [8], while the grain size decreases below 10 nm in the samples produced at lower temperatures or at room temperature. Samples deposited on Si substrates have a lattice parameter $a$ = 535.5 ± 0.5 pm.

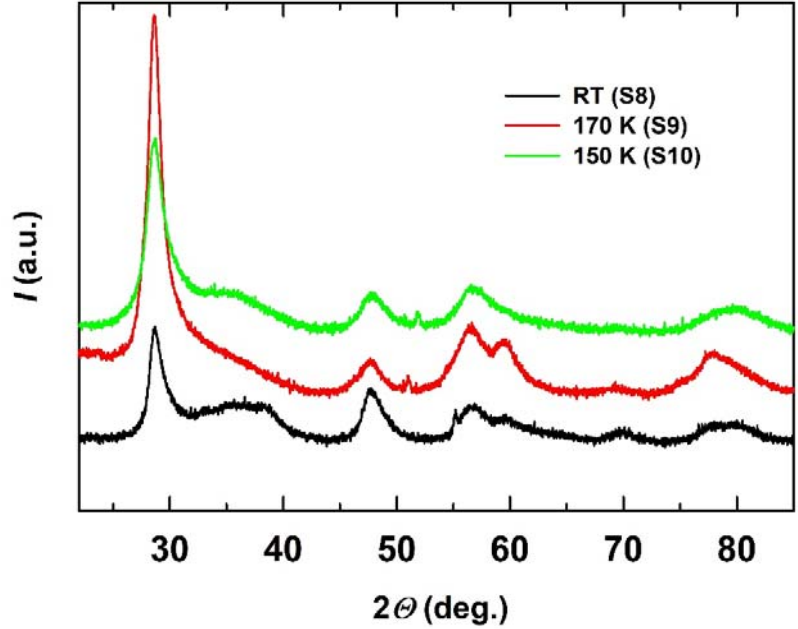

Fig 9. The GIXRD patterns of the U-H films prepared at various temperatures using a Si substrate obtained with an angle of incidence of a primary beam $\alpha_i = 1.5°$.

### 3.3. TEM study of the UH$_2$ film

In order to determine microstructure of the UH$_2$ thin film sample and to obtain elemental mapping, we performed a TEM study of the cross section of the film. Figure 10 shows a global view of the UH$_2$ film on the SiO$_2$ substrate (sample SO6). Location of Pt protection overlayer (used to avoid an excessive damage of the very surface by Ga+ ions during the lamella fabrication), the Mo cap and buffer, and the hydride can be clearly seen in the line profile in Fig. 10(*a*). The total thickness of the thin film in the place where the lamella was extracted from is ≈ 215 nm. The Mo buffer and cap have similar thicknesses, about 9 nm each, they are highlighted with dotted squares in Fig. 10(*a*).

The mapping performed in a scanning TEM mode did not reveal any oxygen in the capped sample, hence we assume the presence of a U hydride only. The issue is to confirm the hydride stoichiometry. The analysis of several regions using the Fast Fourier Transform (FFT) method points to the CaF$_2$ structure type of UH$_2$. One example of the identification of the hydride using the FFT pattern is shown Fig. 10(*b*), in this case, the pattern was generated from the region highlighted by a white circle in the image and corresponds to the [011] axis. When grains are oriented in such way, the {200} and {111} family planes are parallel to the *z* axis, as shown in the FFT picture. The grain size determination performed in several grains indicated an average grain size of about 8 nm (cf. with the refinement of XRD patterns giving 16 nm grain size for this sample (SO6)). The grains are roughly equiaxial, i.e. the film growth is not columnar.

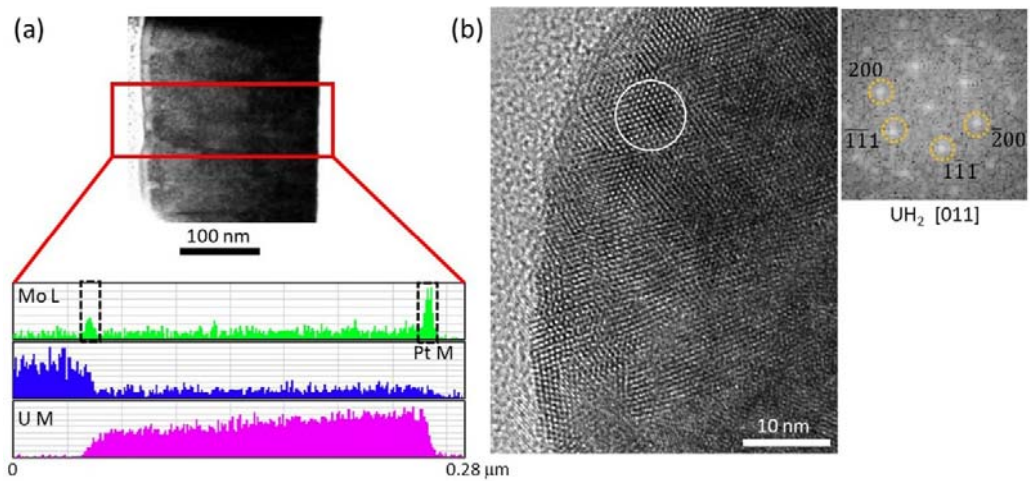

Fig. 10. TEM analysis of the UH$_2$ thin film: line profile showing different elements (a), HRTEM image of the thin film showing the FFT analysis of one grain in the [011] zone axis.

## IV. Physical properties studies

### *4.1. Magnetization study*

Mixture of the UH$_2$ and UH$_3$ phases in some of the samples or presence of the oxidized top surface layer UO$_2$ can be easily recognized using bulk magnetization measurements due to the distinctly different magnetic ordering temperatures. The Curie temperatures $T_C$ determined as inflection points in the temperature dependence of magnetization in low magnetic fields for the two U hydrides, 120 K and 173 K for UH$_2$ and UH$_3$, respectively, allow to distinguish individual increments of spontaneous magnetization. Figure 11 compares zero-field cooling (ZFC) and a field-cooling (FC) magnetization curves measured in the 0.05 T magnetic field applied along the sample surface for various samples. Bulk magnetization of the films is naturally affected by the diamagnetic susceptibility of massive substrates. At fields as low as 0.05 T, the data reflects directly the magnetization while the contribution from the diamagnetic substrate is small. It is seen that all samples are ferromagnetic. In particular, sample SO2 (Fig. 11a) obtained using the lowest current of 0.7 mA, contains both UH$_2$ and UH$_3$ while the increase of the current to 2.2 mA promotes the formation of pure β-UH$_3$ phase. Contamination of the sample SO3 (Fig. 13b) with oxygen explains the small feature in $M(T)$ in the vicinity of the Néel temperature of $UO_2$ ($T_N = 31$ K [18]).

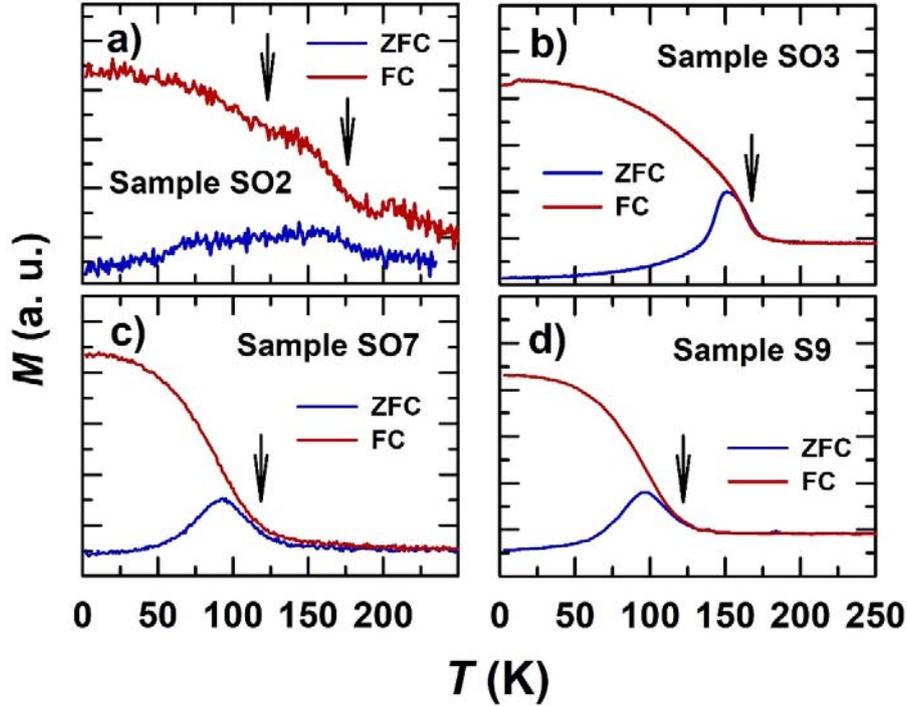

Fig. 11. Temperature dependence of magnetization of the $UH_2$ and $\beta$-$UH_3$ films deposited on various substrates and measured in the field of 0.05 T applied along the sample surface. Arrows indicate ordering temperatures of the phases. ZFC and FC refer to the zero-field-cooled and field-cooled measuring modes.

Collected magnetization data support the XRD analysis conclusion that one route to the $UH_2$ phase is to allow for a mild heating of the substrate (Sample SO7, Fig. 13c). An alternative is to use a Si wafer for substrate (Sample S9, Fig. 11d).

One can see that while the Curie temperature of $UH_2$ is $120 \pm 5$ K, the $\beta$-$UH_3$ films have $T_C$ of $170 \pm 5$ K. For the materials of the $\beta$-$UH_3$ type, the reduced $T_C \approx 145$ K was found only in the case of Mo-diluted $UH_3$ thin film, $(UH_3)_{0.62}Mo_{0.38}$ [6]. (The effect of reduced $T_C$ in Ref. 6 was probably related to the high Mo concentration, dwelling either in the dilution of "magnetic" U atoms or by having the H sites with incomplete occupation. However, we cannot exclude at this level the influence of crystal structure, namely the grains size as small as 1.5 nm in $(UH_3)_{0.62}Mo_{0.38}$.) As pointed out previously [7], small grains size typically leads to the reduction of the ordering temperatures in the U compounds, while the U hydrides in this work, look surprisingly insensitive in this respect, probably because the prominence of the local U-H bonds. Such $T_C$ decrease is opposite to the situation at low Mo alloying, which gives (similar to other transition metal dopants) a systematic increase of $T_C$ up to the maximum at the 12-15% transition metal concentration [19]. The reason can be seen in enhancement of the H/U ratio for Mo concentrations, as Mo occupies the U positions [19], but one cannot exclude even an effect of induced moments on Mo, which vanishes as the U sublattice is progressively incomplete.

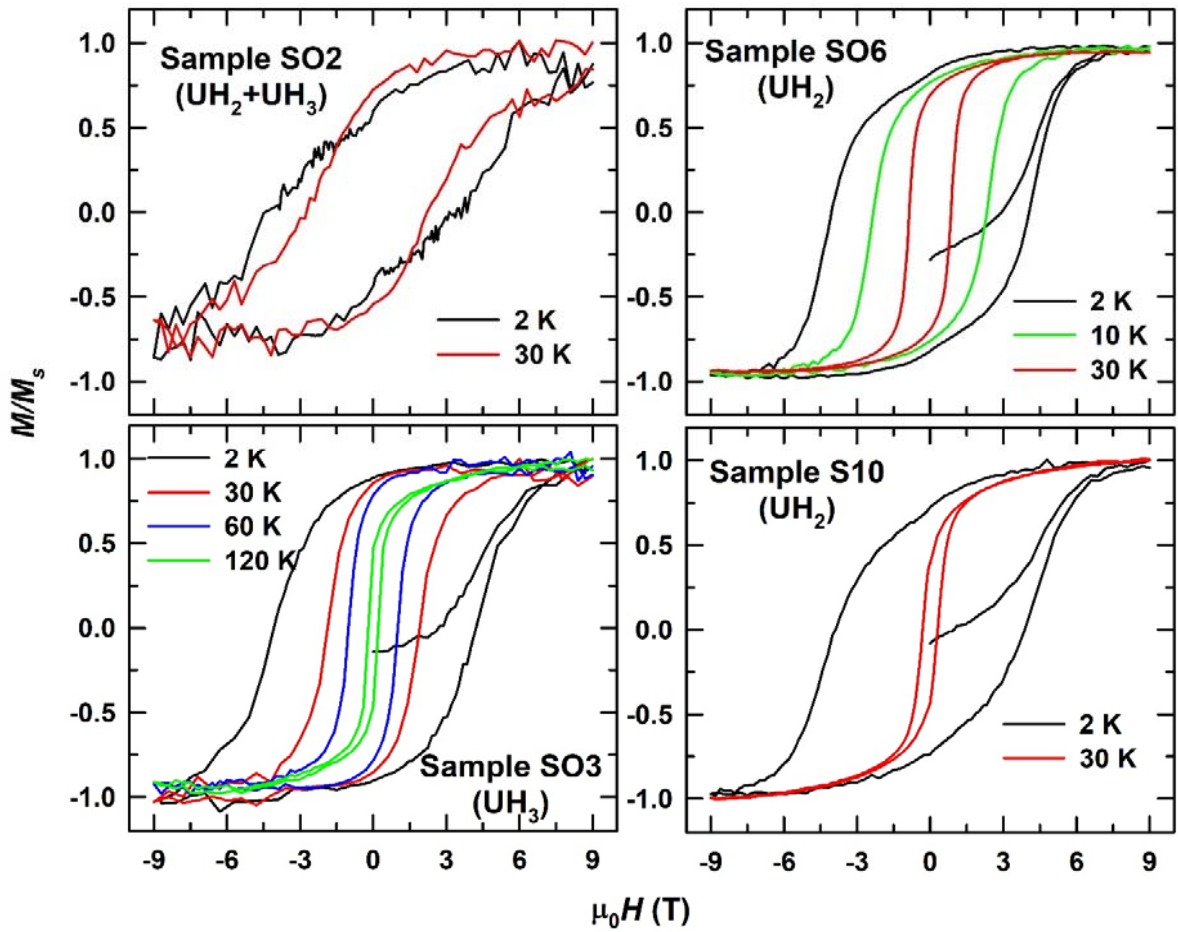

Fig. 12. Magnetization curves of UH$_2$ and β-UH$_3$ samples at various temperatures.

The difference in the ZFC and FC data branches, reflects magnetic history phenomena, given by the hysteresis. The behaviour of the ZFC branch in Fig. 11 indicates that heating nearly to the Curie temperatures is requred to magnetize the samples in the field of $\mu_0 H = 0.05$ T. Such enormous values of hysteresis are related to large magnetic anisotropy, due to sizeable orbital moments inherent to U-based materials. Indeed, all of the $UH_2$ and $UH_3$ samples exhibit very wide magnetic hysteresis loops at low temperatures (see Fig. 12). The hysteresis width decreases with increasing the temperature. Without the precise knowledge of the amount of U in the studied films one cannot determine the size of U moments, therefore we give magnetization in arbitrary units only. An estimate based on the area of the films and approximate thickness gives the order of magnitude, ≈0.5 $\mu_B$/U ($UH_3$ has ≈ 1 $\mu_B$/U [19]). Ab-initio calculations yield the total moment 0.89 $\mu_B$/U [20]. For the majority of samples, the available 9 T field is not strong enough to magnetize the sample, which remains in the regime of minority magnetization loop.

## 4.2. Electrical resistivity

Nearly identical Curie temperatures of various samples obtained at different conditions confirm insensitivity of the U-H magnetism to defects and strains. We further explored the influence of atomic and magnetic disorder on electrical properties of the films. The $UH_3$ films have $\rho(T)$ reminiscent of the bulk $UH_3$ (Fig. 13) with disorder caused by transition metal alloying (T substitutes U atoms) and small grain size [21]. The overall behavior and the absolute values close to 1000 $\mu\Omega$cm are very similar. The cusp is however less sharp. The films, which have no transition metal component, must have however different source of disorder. Besides the small grain size there may be certain randomness in occupation of the H sites.

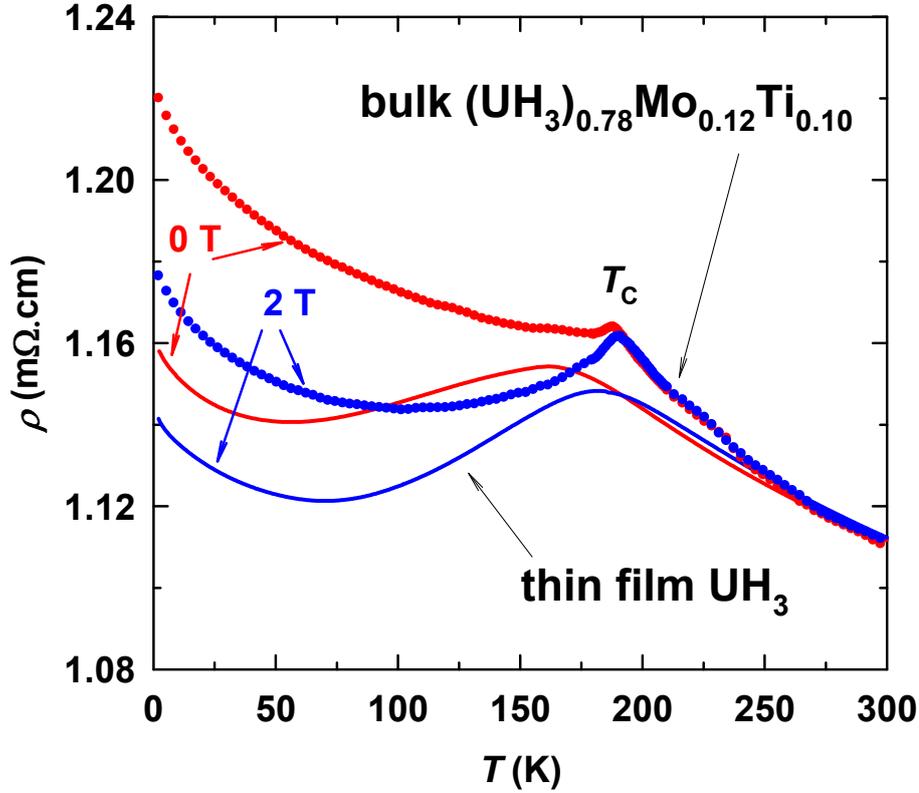

Fig. 13. Comparison of the temperature dependence of electrical resistivity, $\rho(T)$, of the UH$_3$ thin film sample (Sample SO3) with bulk (UH$_3$)$_{0.78}$Mo$_{0.12}$Ti$_{0.10}$ adopted from Ref. [21]. Data in 0 and 2 T fields are shown.

For the UH$_2$ films (Fig. 14), we see that the $T_C$ anomaly is even less pronounced. (The larger the crystallites, the sharper the anomalies associated with the ordering temperature $T_C$.) Similar to UH$_3$, the static magnetic disorder below $T_C$ can be reduced by applied fields, which makes the onset of ferromagnetism more visible. However, the derivative $d\rho/dT$ remains high even below $T_C$, which makes the resistivity increment on cooling from $T = 300$ K reaching up to 17%, i.e. more than double comparing to the UH$_3$ film, in which the magnetic order visibly reduces the increment below $T_C$. The $T_C$ values are in agreement with the magnetization data.

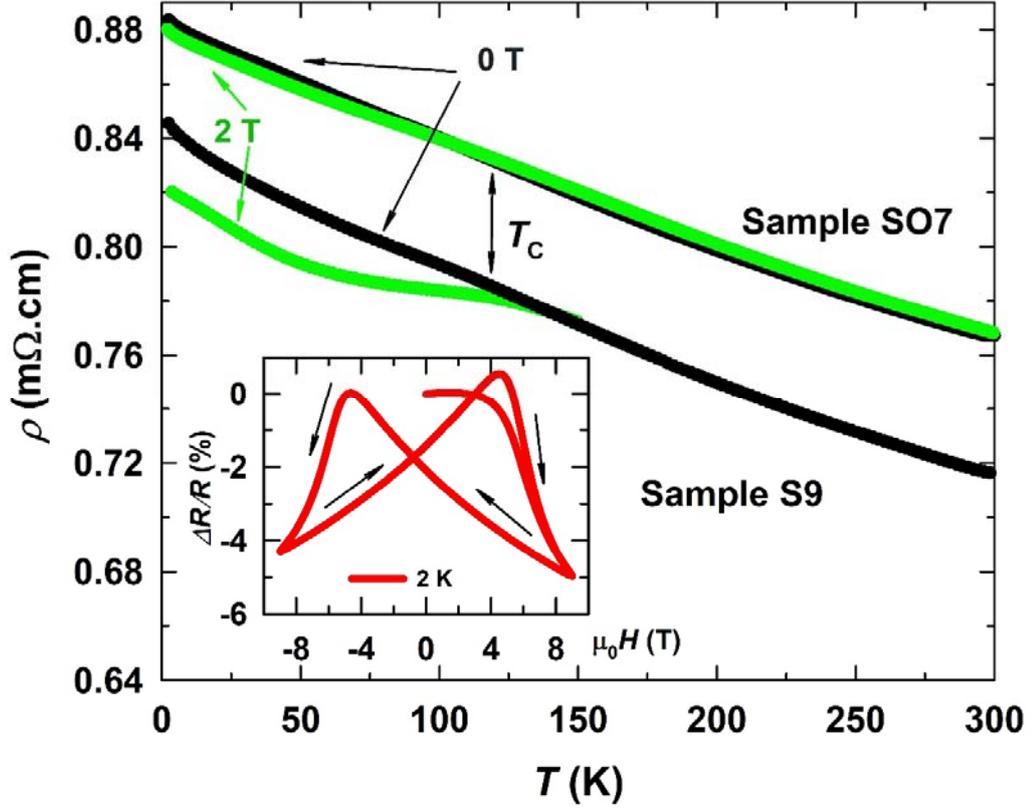

Fig. 14. Comparison of electrical resistivity $\rho(T)$ curves of the $UH_2$ thin film samples grown on a Si substrate (Sample S9) and on $SiO_2$ (self-heated Sample SO7). Data in 0 and 2 T fields are shown. The inset shows the normalized magnetoresistivity data at $T = 2$ K for the Sample S9.

All the materials demonstrate a negative resistivity slope attributed to magnetic disorder, yielding very strong scattering of electrons and weak localization phenomena [21]. Applied magnetic field tends to align individual magnetic moments within nanograins and hence it decreases the resistivity in the ferromagnetic state.

**Discussion and conclusions**

We showed that reactive sputter deposition of U metal can produce clean U hydride films, appearing in two possible forms, depending on sputtering parameters, such as deposition rate or substrate type or temperature. One can obtain either $UH_2$ with the $CaF_2$ structure type (which does not exist in a bulk form) or $UH_3$ with the cubic structure of β-$UH_3$. Both types of films are ferromagnetic. The $T_C$ of the trihydride thin films is close to that in bulk β-$UH_3$ while $T_C$ in $UH_2$ samples grown on two types of substrates under varied temperature conditions and having various grans sizes is ≈ 120 K. This means that unlike most of other U systems, the ferromagnetism is robust and not dependent on structure details on any length scale. However, electric transport does reflect disorder in both cases. This is understandable especially for $UH_2$, because *f*-element hydrides (both lanthanides and actinides) accept H ions entering additional interstitial sites. For rare earths, $REH_{2+x}$ with $x > 0.5$ could be obtained [22]. In $PuH_{2+x}$, $x$ can be as high as 0.7 [17,23]. For β-$UH_3$, the situation is less obvious. The equilibrium phase diagram [24] indicates a possibility of under-stoichiometry ($UH_{3-x}$), but it becomes noticeable only above 600 K. However, non-equilibrium processes connected with sputter deposition can change the situation substantially and the films could be H-deficient. An interesting fact is that no metastable α-$UH_3$ phase has been obtained.

The easiest way to distinguish between the two phases is magnetometry, which allows even to quantify proportions of the phases in mixed-phase films. Performing XRD analysis one has to keep in mind that the films have small grains, but can have also a strong texture.

In future it would be interesting to compare structure features of U-based films prepared as hydrides with those obtained by subsequent hydrogen exposure, as e.g. those described in [25]. Another challenge is to explore possibilities to include alloying with transition metals, known to enhance the $T_C$ values, but possibly making the embedded hydrogen less stable [19].

**Acknowledgements**

The work was supported by the project "Nanomaterials centre for advanced applications", project no. CZ.02.1.01/0.0/0.0/15_003/0000485, financed by ERDF. We also acknowledge the support of Czech Science Foundation under the grant No. 21-09766S. Work of Z.S. was supported in part by the Ministry of Education of the Czech Republic Grants LM2018110 and LNSM-LNSpin. The work at JRC Karlsruhe was performed through ActUsLab/PAMEC under the Framework of access to the Joint Research Centre Physical Research Infrastructures of the European Commission (HYDROFILM_1, AUL-214, RIAA nr Jipsy 35630 and GMR, AUL-213, nr JIPSY 35642). Physical properties measurements were performed in the Materials Growth and Measurement Laboratory (http://mgml.eu/) supported within the program of Czech Research Infrastructures (project no. LM2018096).

**References**

[1]  A. Kaul, O. Gorbenko, M. Novojilov, A. Kamenev, A. Bosak, A. Mikhaylov, O. Boytsova, M. Kartavtseva, Epitaxial stabilization—a tool for synthesis of new thin film oxide materials, J. Cryst. Growth. 275 (2005) e2445–e2451. https://doi.org/10.1016/j.jcrysgro.2004.11.358.

[2]  T. Gouder, R. Eloirdi, F. Wastin, E. Colineau, J. Rebizant, D. Kolberg, F. Huber, Electronic structure of $UH_3$ thin films prepared by sputter deposition, Phys. Rev. B 70 (2004) 1–6. https://doi.org/10.1103/PhysRevB.70.235108.

[3]  R. Troć, W. Suski, The discovery of the ferromagnetism in $U(H, D)_3$: 40 years later, J. Alloys Compd. 219 (1995) 1–5. https://doi.org/10.1016/0925-8388(94)05062-7.

[4]  I. Tkach, M. Paukov, D. Drozdenko, M. Cieslar, B. Vondráčková, Z. Matěj, D. Kriegner, A.V. Andreev, N.-T.H. Kim-Ngan, I. Turek, M. Diviš, L. Havela, Electronic properties of α-UH3 stabilized by Zr, Phys. Rev. B 91 (2015) 115116. https://doi.org/10.1103/PhysRevB.91.115116.

[5]  L. Havela, M. Paukov, M. Dopita, L. Horak, M. Cieslar, D. Drozdenko, P. Minarik, I. Turek, M. Divis, D. Legut, A. Seibert, E. Tereshina-Chitrova, XPS, UPS, and BIS study of pure and alloyed β-$UH_3$ films: Electronic structure, bonding, and magnetism, J. Electron. Spectros. Relat. Phenomena. 239 (2020) 146904. https://doi.org/10.1016/j.elspec.2019.146904.

[6]  E.A. Tereshina-Chitrova, L. Havela, M. Paukov, M. Dopita, L. Horák, O. Koloskova, Z. Šobáň, T. Gouder, F. Huber, A. Seibert, Role of disorder in magnetic and conducting properties of U–Mo and U–Mo–H thin films, Mater. Chem. Phys. 260 (2021) 124069. https://doi.org/10.1016/j.matchemphys.2020.124069.

[7]  L. Havela, K. Miliyanchuk, D. Rafaja, T. Gouder, F. Wastin, Structure and magnetism of thin UX layers, J. Alloys Compd. 408–412 (2006) 1320–1323. https://doi.org/10.1016/j.jallcom.2005.04.125.

[8]  L. Havela, M. Paukov, M. Dopita, L. Horák, D. Drozdenko, M. Diviš, I. Turek, D. Legut, L. Kývala, T. Gouder, A. Seibert, F. Huber, Crystal Structure and Magnetic Properties of Uranium Hydride $UH_2$ Stabilized as a Thin Film, Inorg. Chem. 57 (2018) 14727–14732. https://doi.org/10.1021/acs.inorgchem.8b02499.

[9]  E.A. Tereshina, S. Khmelevskyi, G. Politova, T. Kaminskaya, H. Drulis, I.S. Tereshina, Magnetic ordering temperature of nanocrystalline Gd: Enhancement of magnetic interactions via hydrogenation-induced "negative" pressure, Sci. Rep. 6 (2016) 22553. https://doi.org/10.1038/srep22553.

[10] L. Havela, Hydrogen impact on magnetic properties of metallic systems, J. Alloys Compd. 895 (2022) 162721. https://doi.org/10.1016/j.jallcom.2021.162721.

[11] H.H. Hill, Early actinides: the periodic system's f electron transition metal series, in: W.N. Miner (Ed.), Plutonium 1970 and Other Actinides, 1970: pp. 2–19.

[12] Z. Matěj, A. Kadlecová, M. Janeček, L. Matějová, M. Dopita, R. Kužel, Refining bimodal microstructure of materials with Mstruct, Powder Diffr. 29 (2014) S35–S41. https://doi.org/10.1017/S0885715614000852.

[13] K. Nagao and E. Kagami, Ray Thin-Film Measurement Techniques VII. Pole Figure Measurement, Rigaku Journal 27 (2011) 6–14.

[14] J.W. Ward, L.E. Cox, J.L. Smith, G.R. Stewart, J.H. Wood, Some observations on the electronic structure of β-UD$_3$, Le Journal de Physique Colloques. 40 (1979) C4-15-C4-17. https://doi.org/10.1051/jphyscol:1979403.

[15] O.Yu. Gorbenko, S. v. Samoilenkov, I.E. Graboy, A.R. Kaul, Epitaxial Stabilization of Oxides in Thin Films, Chem. Mater. 14 (2002) 4026–4043. https://doi.org/10.1021/cm021111v.

[16] R.N.R. Mulford, T.A. Wiewandt, The neptunium-hydrogen system, J. Phys. Chem. 69 (1965) 1641–1644. https://doi.org/10.1021/j100889a033.

[17] R.N.R. Mulford, G.E. Sturdy, The Plutonium-Hydrogen System. I. Plutonium Dihydride and Dideuteride, J. American Chem. Soc. 77 (1955) 3449–3452. https://doi.org/10.1021/ja01618a005.

[18] B.T.M. Willis, R.I. Taylor, Neutron diffraction study of antiferromagnetism in $UO_2$, Phys. Letters 17 (1965) 188–190. https://doi.org/10.1016/0031-9163(65)90474-9.

[19] O. Koloskova, V. Buturlim, M. Paukov, P. Minarik, M. Dopita, K. Miliyanchuk, L. Havela, Hydrogen in U-T alloys: Crystal structure and magnetism of $UH_3$-V, J. Alloys Compd. 856 (2021) 157406. https://doi.org/10.1016/j.jallcom.2020.157406.

[20] L. Kývala, L. Havela, A.P. Kadzielawa, D. Legut, Electrons and phonons in uranium hydrides - effects of polar bonding, J. Nuclear Mater. 567 (2022) 153817. https://doi.org/10.1016/j.jnucmat.2022.153817.

[21] L. Havela, M. Paukov, V. Buturlim, I. Tkach, D. Drozdenko, M. Cieslar, S. Mašková, M. Dopita, Z. Matěj, Electrical resistivity of 5f -electron systems affected by static and dynamic spin disorder, Phys. Rev. B 95 (2017) 235112. https://doi.org/10.1103/PhysRevB.95.235112.

[22] P. Vajda, Hydrogen in rare-earth metals, including $REH_{2+x}$ phases, in: K.A. Gschneidner, Jr.L. Eyring (Eds.), Handbook on the Physics and Chemistry of Rare Earths, Elsevier, 1995: pp. 207–291.

[23] R.N.R. Mulford, G.E. Sturdy, The Plutonium-Hydrogen System. II. Solid Solution of Hydrogen in Plutonium Dihydride, J. American Chem. Soc. 78 (1956) 3897–3901. https://doi.org/10.1021/ja01597a010.

[24] I. Grenthe, J. Drożdżynński, T. Fujino, E.C. Buck, T.E. Albrecht-Schmitt, S.F. Wolf, Uranium, in: L.R. Morss, N.M. Edelstein, J. Fuger (Eds.), The Chemistry of the Actinide and Transactinide Elements, Springer Netherlands, Dordrecht, 2006: pp. 253–698. https://doi.org/10.1007/1-4020-3598-5_5.

[25] J.E. Darnbrough, R.M. Harker, I. Griffiths, D. Wermeille, G.H. Lander, R. Springell, Interaction between $U/UO_2$ bilayers and hydrogen studied by in-situ X-ray diffraction, J. Nuclear Mater. 502 (2018) 9–19. https://doi.org/10.1016/j.jnucmat.2018.01.031.